\documentclass[prl,twocolumn,showpacs,superscriptaddress,amsmath]{revtex4}

\usepackage{bm}
\usepackage{graphicx}

\newcommand{\bra}[1]{\langle #1 | \,}
\newcommand{\ket}[1]{\, | #1 \rangle}

\newcommand{\expv}[1]{\langle #1 \rangle}
\newcommand{\ddt}{\frac{\partial}{\partial t}}

\newcommand{\la}{\lambda}
\newcommand{\om}{\omega}
\newcommand{\Om}{\Omega}
\newcommand{\ga}{\gamma}

\newcommand{\de}{\delta}
\newcommand{\De}{\Delta}
\newcommand{\ka}{\kappa}
\newcommand{\eps}{\epsilon}
\newcommand{\br}{\mathbf{r}}
\newcommand{\Eh}{\hat{\cal E}}
\newcommand{\Psih}{\hat{\Psi}}

\newcommand{\alh}{\hat{\alpha}}
\newcommand{\Ih}{\hat{\cal I}}
\newcommand{\sih}{\hat{\sigma}}
\newcommand{\zp}{z^{\prime}}
\newcommand{\tp}{t^{\prime}}

\begin{document}

\title{Long-Range Interactions and Entanglement of Slow Single-Photon Pulses}

\author{Inbal Friedler}
\affiliation{Department of Chemical Physics, 
Weizmann Institute of Science, Rehovot 76100, Israel}
\author{David Petrosyan}
\affiliation{Institute of Electronic Structure \& Laser, 
FORTH, GR-71110 Heraklion, Crete, Greece}
\author{Michael Fleischhauer}
\affiliation{Fachbereich Physik, Universit\"at Kaiserslautern, 
D-67663 Kaiserslautern, Germany}
\author{Gershon Kurizki}
\affiliation{Department of Chemical Physics, 
Weizmann Institute of Science, Rehovot 76100, Israel}

\date{\today}

\begin{abstract}
We show that very large nonlocal nonlinear interactions between 
pairs of colliding slow-light pulses can be realized in atomic 
vapors in the regime of electromagnetically induced transparency. 
These nonlinearities are mediated by strong, long-range dipole--dipole 
interactions between Rydberg states of the multi-level atoms in a 
ladder configuration. In contrast to previously studied schemes, this 
mechanism can yield a homogeneous conditional phase shift of $\pi$ even 
for weakly focused single-photon pulses, thereby allowing a deterministic
realization of the photonic phase gate.
\end{abstract}


\maketitle


Whether or not quantum information processing and quantum computing 
\cite{QCcomp} become practical technologies crucially depends on the 
ability to implement high-fidelity quantum logic gates in a scalable way 
\cite{diVinc}. Among alternative routes to this challenging goal, are of 
particular interest the schemes operating with photons as qubits 
\cite{photqc,linopt}, since photons are ideal carriers of quantum 
information in terms of transfer rates, distances and scalability. 
A current trend makes use of linear optical elements and photodetectors 
for the implementation of key components of quantum communications and 
information processing in a probabilistic way \cite{linopt}. The desirable
objective though is a {\em deterministic} realization of entangling 
operations between individual photons, which require sufficiently strong 
nonlinearities or long interaction times. These are achievable, 
at the single-photon level, by tight spatial confinement of the photons, 
in the very demanding regime of strong atom-field coupling in high-$Q$ 
cavities \cite{phphcav}.

A promising alternative is to enhance both the nonlinear susceptibility 
and interaction time, by employing the ultra-slow light propagation in 
resonant media subject to electromagnetically induced transparency (EIT)
\cite{EIT,ScZub,vred}. In a pioneering work, Schmidt and Imamo\u{g}lu have 
suggested the possibility of enhanced, non-absorptive, cross-phase modulation
of two weak fields in the EIT regime \cite{imam}, provided their interaction 
time is long enough. However, upon entering the EIT medium light pulses
become spatially compressed by the ratio of group velocity $v$ to 
the vacuum speed of light $c$ \cite{harhau}, so that the interaction time 
of two colliding pulses is a constant independent of $v$. In order to maximize
this time, copropagating pulses with nearly matched group velocities have 
been proposed \cite{lukimam,petmal}. The essential drawback of such an 
approach is the spatial inhomogeneity of the conditional phase shift, causing
spectral broadening of the interacting pulses, thereby preventing the 
realization of a high-fidelity quantum phase gate. Alternative approaches
free of spectral broadening have been suggested \cite{IFGKDP,lukin-pbg,MMMF}.
In all of them, however, a rather tight transverse confinement through
waveguiding or focusing of the pulses, close to the diffraction limit of 
$\la^2$, is needed in order to attain a phase shift of $\pi$, which is 
technically challenging. 

When the light pulses enter EIT media, photonic excitations are temporarily
transferred to atomic excitations through the formation of quasi-particles, 
the so-called dark-state (or slow-light) polaritons, which are superpositions
of light and matter degrees of freedom \cite{fllk}. The spatial compression 
of the pulses leads to an {\em amplification} of the matter components of 
polaritons. In this Letter we propose a hitherto unexplored mechanism for the 
collisional entanglement of two single-quantum polaritons mediated by 
the long-range interaction of their matter (atomic) components and 
demonstrate its effectiveness. In contrast to the previous schemes which 
employ {\em local} interactions, namely either two photons interact with 
the same atom \cite{lukimam,petmal,IFGKDP,lukin-pbg} or two atoms after
absorbing the photons undergo $s$-wave scattering \cite{MMMF}, here the
two polaritons interact via the long-range dipole-dipole interactions
between their atomic components in the highly excited Rydberg states.
In a static electric field, these internal Rydberg states, populated 
only in the presence of polaritons, possess large permanent dipole 
moments \cite{RydAtoms}, which can further enhance the effective 
interaction time between the polaritons. We will show that under 
experimentally realizable conditions, the conditional phase shift 
accumulated during a collision of two single-quantum polaritons 
is {\em spatially homogeneous} and can be sufficiently large for 
the implementation of the quantum phase gate, even for moderate 
focusing or transverse confinement of interacting pulses. 
We note that quantum gates for individual Rydberg atoms, coupled by 
dipole-dipole interaction, has been proposed in \cite{JCZRCL}.


\begin{figure}[t]
\includegraphics[width=8.5cm]{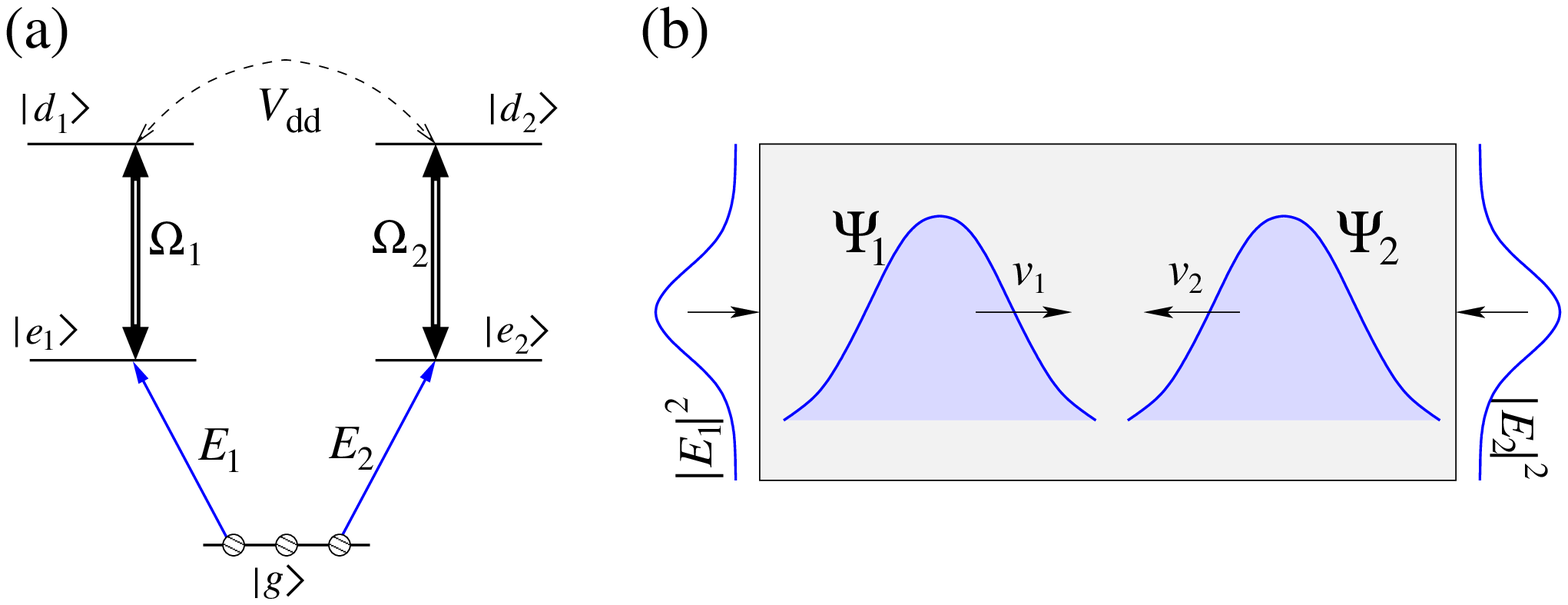}
\caption{(a)~Level scheme of atoms interacting with weak (quantum) 
fields $E_{1,2}$ on the transitions $\ket{g} \to \ket{e_{1,2}}$ 
and strong driving fields of Rabi frequencies $\Om_{1,2}$ on 
the transitions $\ket{e_{1,2}} \to \ket{d_{1,2}}$, respectively. 
$V_{\rm dd}$ denotes the dipole-dipole interaction between pairs 
of atoms in Rydberg states $\ket{d}$.
(b)~Upon entering the medium, each field having Gaussian transverse
intensity profile is converted into the corresponding polariton 
$\Psi_{1,2}$ representing a coupled excitation of the field and 
atomic coherence. These polaritons propagate in the opposite directions 
with slow group velocities $v_{1,2}$ and interact via the dipole-dipole 
interaction.} 
\label{fig:als}
\end{figure}

We consider an ensemble of cold alkali atoms with level configuration 
as in Fig.~\ref{fig:als}. All the atoms are initially prepared
in the ground state $\ket{g}$. Two weak (quantum) fields $E_{1,2}$ 
having orthogonal polarizations and propagating in the opposite 
directions along the $z$ axis resonantly interact with
the atoms on the transitions $\ket{g} \to \ket{e_{1,2}}$, respectively. 
The intermediate states $\ket{e_{1,2}}$ are resonantly coupled by two 
strong (classical) driving fields with Rabi frequencies $\Omega_{1,2}$ 
to the highly excited Rydberg states $\ket{d_{1,2}}$. In a static electric
field $E_{\rm st} \mathbf{e}_z$, the Rydberg states $\ket{d}$ possess large 
permanent dipole moments $\mathbf{p}= \frac{3}{2} n q e a_0 \mathbf{e}_z$,
where $n$ and $q \equiv n_1 - n_2$ are, respectively, the (effective) 
principal and parabolic quantum numbers, $e$ is the electron charge, and 
$a_0$ is the Bohr radius \cite{RydAtoms}. A pair of atoms $i$ and $j$ at 
positions $\br_i$ and $\br_j$ excited to states $\ket{d}$ interact 
with each other via the dipole-dipole potential
\[
V_{\rm dd} = \frac{\mathbf{p}_i \cdot \mathbf{p}_j 
- 3 (\mathbf{p}_i \cdot \mathbf{e}_{ij}) (\mathbf{p}_j \cdot \mathbf{e}_{ij})}
{4 \pi \eps_0 |\br_i -\br_j|^3} , 
\]
where $\mathbf{e}_{ij}$ is a unit vector along the interatomic direction.
This dipole-dipole interaction results in an energy shift of the pair of 
Rydberg atoms, while we assume that the state mixing within the same $n$ 
manifold is suppressed by the proper choice of parabolic $q$ and magnetic
$m$ quantum numbers \cite{RydAtoms,JCZRCL}. In the frame rotating with the 
frequencies of the optical fields, the interaction Hamiltonian has the 
following form
\begin{equation}
H = V_{\rm af} + V_{\rm dd} ,
\end{equation}
where the atom-field and dipole-dipole interaction terms are given, 
respectively, by 
\begin{subequations}
\label{VafVdd}
\begin{eqnarray}
V_{\rm af} &=& - \hbar \sum_j^N [g_1^j \Eh_1 \sih_{e_1 g}^j 
+ \Om_1 \sih_{d_1 e_1}^j 
\nonumber \\ & & \;\;\;\;\;\;\;\;\;\;
+g_2^j \Eh_2 \sih_{e_2 g}^j + \Om_2 \sih_{d_2 e_2}^j 
+ {\rm H. c.}], \\
V_{\rm dd} &=& \hbar \sum_{i > j}^N 
\sih_{d d}^i \De(\br_i -\br_j) \sih_{d d}^j .
\end{eqnarray}
\end{subequations}
Here $N = \rho V$ is the total number of atoms, $\rho$ being the (uniform) 
atomic density and $V$ the volume; 
$\sih_{\mu \nu}^j \equiv \ket{\mu}_{jj}\bra{\nu}$
is the transition operator of the $j$th atom; $\Eh_l$ is the 
slowly-varying operator, corresponding to the electric field $E_l$ ($l=1,2$), 
which obeys the commutation relations 
$[\Eh_l(\br),\Eh^{\dagger}_{l^{\prime}}(\br^{\prime})] 
= V \de_{l l^{\prime}} \de(\br - \br^{\prime})$;
$g_l^j$ is the corresponding atom-field coupling constant on the transition 
$\ket{g}_j \to \ket{e_l}_j$; and $\hbar \De(\br_i -\br_j) \equiv 
\, _i \bra{d} _j \bra{d} V_{\rm dd} \ket{d}_i \ket{d}_j$
is the dipole-dipole energy shift for a pair of atoms $i$ and $j$, 
given by
\[
\De(\br_i -\br_j) = C \, \frac{1 - 3 \cos^2 \vartheta}{|\br_i -\br_j|^3} ,
\] 
where $\vartheta$ is the angle between vectors $\mathbf{e}_z$ and 
$\mathbf{e}_{ij}$, 
and $C = \wp_{d_l} \wp_{d_{l^{\prime}}}/(4 \pi \eps_0 \hbar)$
is a constant proportional to the product of atomic dipole moments 
$\wp_{d_l} = \bra{d_l} \mathbf{p} \ket{d_l}$ assumed the same for both 
states $\ket{d_{1,2}}$, $\wp_{d_{1,2}} = \wp_d$. 

Let us introduce collective atomic operators
$\sih_{\mu \nu}(\br) = \frac{1}{N_r} \sum_{j=1}^{N_r} \sih_{\mu \nu}^j$ 
averaged over the volume element $d^3 r$ containing $N_r = \rho \, d^3 r \gg 1$
atoms around position $\br$. Then Eqs.~(\ref{VafVdd}) can be cast in the 
continuous form
\begin{subequations}
\label{VVcont}
\begin{eqnarray}
V_{\rm af} &=& - \hbar \rho \int d^3 r \sum_{l=1,2} 
[g_l \Eh_l  \sih_{e_l g}(\br) 
+ \Om_l \sih_{e_l d_l}(\br)]
+ {\rm H. c.} ,\;\;\;\;\;  \\
V_{\rm dd} &=& \hbar \rho^2 \int \! \! \! \int d^3 r \, d^3 r^{\prime}
\sih_{d d}(\br) \De(\br -\br^{\prime}) \sih_{d d} (\br^{\prime}) .
\end{eqnarray}
\end{subequations}
Using Eqs.~(\ref{VVcont}), one can derive a set of Heisenberg-Langevin
equations for the atomic operators $\sih_{\mu \nu}$ \cite{ScZub}. 
When the number of photons in the quantum fields $\Eh_l$ is much smaller 
than the number of atoms, these equations can be solved perturbatively 
in the small parameters $g_l \Eh_l/\Om_l$ and in the adiabatic 
approximation for all the fields \cite{fllk}, with the result
\begin{subequations}
\label{sigmas}
\begin{eqnarray}
\sih_{ge_l}(\br) &=& -\frac{i}{\Om_l} 
\left[ \ddt + i \alh(\br) \right] \sih_{gd_l}(\br) , \\
\alh(\br) &= & \rho \int d^3 r^{\prime} \De(\br -\br^{\prime}) 
[\sih_{d_1 d_1}(\br^{\prime}) + \sih_{d_2 d_2}(\br^{\prime})] , \quad \\
\sih_{gd_l}(\br) &=& - \frac{g_l \Eh_l}{\Om_l^*} , \;\;\;\;
\sih_{d_l d_l}(\br) = \sih_{d_l g}(\br) \sih_{gd_l}(\br) .
\end{eqnarray}
\end{subequations}
Let us assume that the transverse profile of both quantum fields is 
described by a Gaussian $e^{-r_{\bot}^2/w^2}$  of width $w$, where 
$r_{\bot} = |\br_{\bot}|$ is the distance from the field propagation
axis, while the Rabi frequencies of classical driving fields $\Om_l$ 
are uniform over the entire volume $V$. We may then write 
$g_l \Eh_l = g_l(\br_{\bot}) \Eh_l(z)$, where the traveling-wave electric 
field operators $\Eh_l(z) = \sum_k a_l^k e^{ikz}$ are expressed through 
the superposition of bosonic operators $a_l^k$ for the longitudinal field 
modes $k$, while the (transverse-position-dependent) coupling constants 
are given by $g_l(\br_{\bot}) = \tilde{g}_l e^{-r_{\bot}^2/2 w^2}$, with 
$\tilde{g}_l = (\wp_{ge_l}/\hbar) \sqrt{\hbar \om/2 \eps_0 V}$, $\wp_{ge_l}$
being the dipole matrix element on the transition $\ket{g} \to \ket{e_l}$,
$V = \pi w^2 L$, and $L$ the medium length. Under this approximation, the 
propagation equations for the slowly-varying quantum fields have the form
\begin{equation}
\left(\frac{\partial}{\partial t} \pm c\frac{\partial}{\partial z}\right) 
\Eh_l(z,t) = i \tilde{g}_l N \sih_{g e_l}(z), \label{Eprop} 
\end{equation}
the sign ``$+$'' or ``$-$'' corresponding to $l = 1$ or $2$, respectively.

\begin{figure}[t]
\includegraphics[width=7cm]{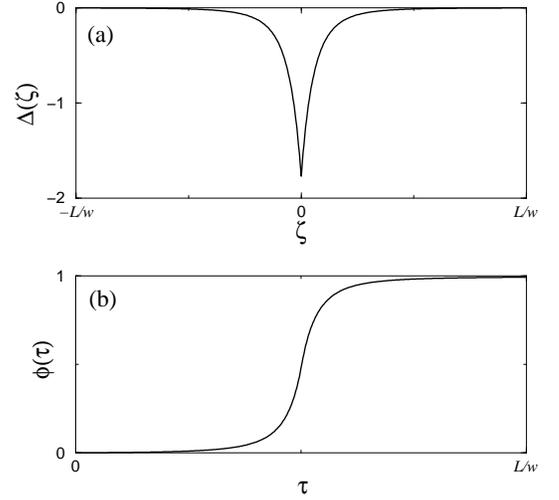}
\caption{(a)~The 1D dipole-dipole potential $\De({\zeta})$ of 
Eq.~(\ref{1Dddpot}) as a function of dimensionless distance 
$\zeta = (z - \zp)/w$, in units of $2 C/w^3$ Hz.
(b)~ The resulting phase-shift $\phi(\tau) \equiv \phi(vt,L-vt,t)$
of Eq.~(\ref{phiphsh}) as a function of dimensionless time 
$\tau = vt/w$, in units of $2 C/(v w^2)$ rad.} 
\label{fig:phshgr}
\end{figure}

Following \cite{fllk}, we introduce new quantum fields $\Psih_l$---dark
state polaritons---via the canonical transformations
\begin{equation}
\Psih_l = \cos \theta_l \Eh_l - \sin \theta_l \sqrt{N} \sih_{gd_l} , 
\label{polars}
\end{equation}
where the mixing angles $\theta_l$ are defined through 
$\tan^2 \theta_l = \tilde{g}_l^2 N/|\Om_l|^2$. These polaritons
correspond to coherent superpositions of electric field $\Eh_l$ 
and atomic coherence $\sih_{gd_l}$ operators. Employing the 
plane-wave decomposition of the polariton operators, one can 
show that in the weak-field limit, they obey the bosonic commutation
relations $[\Psih_{l}(z),\Psih_{l^{\prime}}^{\dagger}(\zp)] 
\simeq L\de_{ll^{\prime}} \de(z-\zp)$. Using Eqs.~(\ref{sigmas}) 
and (\ref{Eprop}), we obtain the following propagation equations for 
the polariton operators,
\begin{equation}
\left(\frac{\partial}{\partial t} \pm v_l\frac{\partial}{\partial z}\right) 
\Psih_l(z,t) = - i \sin^2 \theta_l \alh(z,t)\Psih_l(z,t) . \label{Psiprop} 
\end{equation}
Here $v_l = c \cos^2 \theta_l$ is the group velocity, while operator 
$\alh(z,t)$ is responsible for the self- and cross-phase modulation between 
the polaritons. It is related to the polariton intensity (excitation number)
operators $\Ih_l \equiv \Psih_l^{\dagger} \Psih_l$ via
\begin{equation}
\alh(z,t) = \frac{1}{L} \int_0^L \!\! d \zp \De(z - \zp) 
[\sin^2 \theta_1 \Ih_1(\zp, t) + \sin^2 \theta_2 \Ih_2(\zp, t)] ,
\end{equation}
where the 1D dipole-dipole interaction potential $\De(z - \zp)$ is obtained
after the integration over the transverse profile of the quantum fields,
\begin{eqnarray}
\De(z - \zp) &=& \frac{1}{\pi w^2} \int_{0}^{2 \pi} \!\! d \varphi^{\prime}  
\!\! \int_{0}^{\infty} \!\! d r^{\prime}_{\bot} r^{\prime}_{\bot} 
e^{-r^{\prime 2}_{\bot}/w^2} \De(z \mathbf{e}_z - \br^{\prime}) \nonumber \\ 
&=& \frac{2 C}{w^3}
\left[ \frac{2 |z -\zp|}{w} 
-\sqrt{\pi} \left(1+ 2 \frac{|z -\zp|^2}{w^2} \right) 
\right. \nonumber \\ & & \;\;\;\; \left. \times 
\exp \left( \frac{|z -\zp|^2}{w^2} \right)
{\rm erfc}\left( \frac{|z -\zp|}{w} \right) \right] , \label{1Dddpot}
\end{eqnarray}
and is shown in Fig.~\ref{fig:phshgr}(a).

It follows from Eq.~(\ref{Psiprop}) that the intensity operators 
$\Ih_l$ are constants of motion: $\Ih_l(z,t) = \Ih_l(z \mp v_l t,0)$, 
the upper (lower) sign corresponding to $l=1$ ($l=2$). Then the formal 
solution for the polariton operators can be written as
\begin{eqnarray}
\Psih_l(z,t) & = & 
\exp \left[- i \sin^2 \theta_l \int_0^t \!\! d \tp 
\alh(z \mp v_l(t-\tp),\tp) \right]
\nonumber \\ & & \;\; \times 
\Psih_l(z \mp v_l t,0) . \label{Psisolv}
\end{eqnarray}

Equation (\ref{Psisolv}) is our central result. Let us outline the 
approximations involved in the derivation of this solution. In order 
to accommodate the pulses in the medium with negligible losses, their 
duration $T$ should exceed the inverse of the EIT bandwidth
$\de \om = |\Om_l|^2 (\ga_{ge_l} \sqrt{\ka_0 L})^{-1}$,
where $\ga_{ge_l}$ is the transversal relaxation rate and 
$\ka_0 \simeq 3 \la^2 /(2 \pi)\rho$ is the resonant absorption coefficient 
on the transition $\ket{g} \to \ket{e_l}$. This yields the condition
$(\ka_0 L)^{-1/2} \ll T v_l/L < 1$ which requires a medium with large 
optical depth $\ka_0 L \gg 1$ \cite{fllk}. In addition, the dipole-dipole
energy shift should lie within the EIT bandwidth $\de \om$ for all 
$|z-\zp| \leq L$, which implies that $|\De(0)|=2 \sqrt{\pi} C/w^3 < \de \om$. 
Finally, the propagation/interaction time of the two pulses 
$t_{\rm out} = L/v_l$ is limited by the relaxation rate of the 
Rydberg states $\ga_{d_l}$ via $t_{\rm out} \ga_{d_l} \ll 1$.


From now on, we assume that $\theta_{1,2} = \theta$, i.e.,
$\tilde{g}_1^2 N/|\Om_1|^2 = \tilde{g}_2^2 N/|\Om_2|^2$, 
which yields $v_{1,2} = v = c \cos^2 \theta$.
We are interested in the evolution of input state 
\begin{equation}
\ket{\Phi_{\rm in}} = \ket{1_1} \otimes \ket{1_2} ,
\end{equation}
composed of two single-excitation polariton wavepackets 
\[
\ket{1_l} = \frac{1}{L} \int \! dz f_l(z) \Psih_l(z)^{\dagger} \ket{0},
\]
where $f_l(z)$ define the spatial envelopes of the corresponding wavepackets 
$l=1,2$ which initially (at $t=0$) are localized around $z=0,L$, respectively.
For such an initial state, all the relevant information is contained in the 
expectation values of the polariton intensities 
$\expv{\Ih_l(z,t)} = \bra{\Phi_{\rm in}} \Ih_l(z,t) \ket{\Phi_{\rm in}}$
and the two-particle wavefunction \cite{ScZub,lukimam,petmal}
\begin{equation}
F_{12}(z_1,z_2,t) = \bra{0} \Psih_1(z_1,t) \Psih_2(z_2,t)\ket{\Phi_{\rm in}} 
\label{tpwv}. 
\end{equation}
With the above solution, for the polariton intensities we have
$\expv{\Ih_{1,2}(z,t)}=\expv{\Ih_{1,2}(z \mp vt,0)}=|f_{1,2}(z \mp vt)|^2$, 
which describes the shape-preserving counter-propagation of the two polaritons
with group velocity $v$. Substituting the operator solution (\ref{Psisolv}) 
into (\ref{tpwv}), after some algebra, we obtain the following expression 
for the two-particle wavefunction
\begin{eqnarray}
F_{12}(z_1,z_2,t) &=& f_1(z_1 - vt) f_2(z_2 + vt) \exp[i \phi(z_1,z_2,t)] , 
\qquad \\
\phi(z_1,z_2,t) &=& - \sin^4 \theta \int_0^t \!\! d \tp 
\De(z_1 - z_2 - 2 v (t - \tp) ) , \label{phiphsh}
\end{eqnarray}
which indicates that the dipole-dipole interaction between the two
single-excitation polaritons results in the conditional phase-shift
$\phi(z_1,z_2,t)$. We consider a situation in which at time $t=0$,
the first pulse is localized at $z_1 =0$ and the second pulse is at
$z_2 = L$, while after the interaction, at time $t_{\rm out} = L/v$, 
the coordinates of the two pulses are $z_1 = L$ and $z_2 = 0$, respectively 
[Fig.~\ref{fig:phshgr}(b)]. Then the phase-shift accumulated during the 
interaction is spatially uniform, and is given by
\begin{equation}
\phi(L,0,L/v) = - \frac{\sin^4\theta}{v} \int_0^L \!\! d \zp 
\De(2 \zp -L ) = \frac{2 C \sin^4 \theta}{v w^2} .
\end{equation}
This remarkably simple result is obtained upon replacing the variable 
$(2 \zp -L)/w \to \zeta^{\prime}$ and extending the integration limits 
to $L/w \to \infty$. The main limitation on the phase shift is imposed 
by the condition $|\De(0)| < \de \om$. In terms of experimentally relevant
parameters, the group velocity is $v \simeq 2 |\Om|^2 /(\ka_0 \ga_{ge}) \ll c$
($\sin^2 \theta \simeq 1$), and we have $\phi <\frac{1}{2}w\sqrt{\ka_0/\pi L}$.

To relate the foregoing discussion to a realistic experiment, let us 
assume an ensemble of cold alkali atoms in the ground state $\ket{g}$ 
with density $\rho \sim 10^{14}$~cm$^{-3}$ confined in a trap of length 
$L \sim 100 \;\mu$m. The resonant quantum fields with $\la \sim 0.5\;\mu$m 
have the transverse width $w \sim 30\;\mu$m. In the presence of driving
fields with appropriate frequencies, the single-photon pulses lead to the 
(two-photon) excitation of single atoms to the Rydberg states $\ket{d}$ 
with quantum numbers $n \simeq 25$ and $q=n-1$. The corresponding dipole 
moments are $\wp_d \simeq 900 e a_0$, while 
$\ga_d \sim 2 \times 10^3$~s$^{-1}$ \cite{RydAtoms}. With 
$\ga_{ge} \sim 10^7$~s$^{-1}$ and $\Om \sim 1.6 \times 10^7$~rad/s, 
the group velocity is $v \simeq 4$~m/s, and the accumulated phase shift
is $\phi \simeq \pi$ with the fidelity $F = \exp(-\ga_d L/v)\gtrsim 0.95$.


To summarize, we have studied a novel highly-efficient scheme for 
cross-phase modulation and entanglement of two counterpropagating 
single-photon wavepackets, employing their ultra-small group velocities 
in atomic vapors, under the conditions of electromagnetically induced 
transparency, and the strong long-range dipole-dipole interactions of
the accompanying Rydberg-state excitations in a ladder-type field-atom
coupling setup. We have solved,  in the weak-field and adiabatic 
approximations, the effective one-dimensional propagation equations 
for the polariton operators and have shown that the dipole-dipole 
interaction leads to a {\em homogeneous} conditional phase shift
that reach the value of $\pi$ even if the transverse cross section of 
the pulses $w^2$ is much (three orders of magnitude) larger than the 
diffraction limit $\la^2$. This is the obvious merit of the present
proposal, as compared to previous schemes based on local interactions 
of photons or slow-light polaritons 
\cite{imam,harhau,lukimam,petmal,IFGKDP,lukin-pbg,MMMF},
which require the photonic beam cross section to be comparable to the cross
section for atomic resonant absorption. Hence, our proposal paves the
way to the coveted deterministic entanglement of two single-photon pulses
and the realization of the universal photonic phase gate \cite{IFGKDP}.

\begin{acknowledgments}
This work was supported by the EC (QUACS RTN and ATESIT network), 
ISF, and Minerva.
\end{acknowledgments}


\begin{thebibliography}{99}

\bibitem{QCcomp} M.~Nielsen and I.~Chuang,
{\it Quantum Computation and Quantum Information}, 
(Cambridge University Press, Cambridge, 2000).

\bibitem{diVinc} D.P.~DiVincenzo, Fortschr. Phys. {\bf 48}, 711 (2000).

\bibitem{photqc}
I.L.~Chuang and Y.~Yamamoto, Phys. Rev. A {\bf 52}, 3489 (1995).

\bibitem{linopt}
E.~Knill, R.~Laflamme, and G.J.~Milburn, Nature {\bf 409}, 46 (2001);
J.L.~O'Brien {\it et al.}, Nature {\bf 426}, 264 (2003);
S.~Gasparoni {\it et al.}, Phys. Rev. Lett. {\bf 93}, 020504 (2004).

\bibitem{phphcav}
Q.~A. Turchette {\it et al.}, Phys. Rev. Lett. {\bf 75}, 4710 (1995);
A.~Imamoglu, H.~Schmidt, G.~Woods, and M.~Deutsch, 
Phys. Rev. Lett. {\bf 79}, 1467 (1997); 
A.~Rauschenbeutel {\it et al.}, Phys. Rev. Lett. {\bf 83}, 5166 (1999).

\bibitem{EIT} S. E. Harris, Phys. Today {\bf 50}(7), 36 (1997);
M. D. Lukin, Rev. Mod. Phys. {\bf 75}, 457 (2003).

\bibitem{ScZub}
M.O.~Scully and M.S.~Zubairy, {\em Quantum Optics}
(Cambridge University Press, Cambridge, UK, 1997).

\bibitem{vred} 
L.V.~Hau, S.E.~Harris, Z.~Dutton, and C.H.~Behroozi,
Nature {\bf 397}, 594 (1999);
M.M.~Kash {\it et al.}, Phys. Rev. Lett. {\bf 82}, 5229 (1999); 
D.~Budker, D.F.~Kimball, S.M.~Rochester, and V.V.~Yashchuk, 
Phys. Rev. Lett. {\bf 83}, 1767 (1999).

\bibitem{imam}
H.~Schmidt and A.~Imamo\u{g}lu, Opt. Lett. {\bf 21}, 1936 (1996).

\bibitem{harhau}
S.~Harris and L.~Hau, Phys. Rev. Lett. {\bf 82}, 4611 (1999).

\bibitem{lukimam}
M.D.~Lukin and A.~Imamo\u{g}lu, Phys. Rev. Lett. {\bf 84}, 1419 (2000).

\bibitem{petmal}
D. Petrosyan and Yu. P. Malakyan, Phys. Rev. A {\bf 70}, 023822 (2004).

\bibitem{IFGKDP} 
I. Friedler, G. Kurizki and D. Petrosyan,  
Europhys. Lett. {\bf 68}, 625 (2004); 
Phys. Rev. A {\bf 71}, 023803 (2005).

\bibitem{lukin-pbg}
A.~Andre M.~Bajcsy, A.S.~Zibrov, and M.D.~Lukin, 
Phys. Rev. Lett. {\bf 94}, 063902 (2005).

\bibitem{MMMF} 
M. Masalas and M. Fleischhauer, Phys. Rev. A {\bf 69}, 061801(R) (2004).

\bibitem{fllk}
M.~Fleischhauer and M.D.~Lukin, Phys. Rev. Lett. {\bf 84}, 5094 (2000);
Phys. Rev. A {\bf 65}, 022314 (2002).

\bibitem{RydAtoms}
T.F.~Gallagher, {\em Rydberg Atoms} (Cambridge University Press, 
Cambridge, 1994).

\bibitem{JCZRCL} 
D.~Jaksch {\it et al.}, Phys. Rev. Lett. {\bf 85}, 2208 (2000).


\end{thebibliography}
\end{document}